\documentclass{emulateapj}

\usepackage{float,epsfig}

\usepackage{pstricks}

 \shorttitle{X-ray galaxy groups in CNOC2}

 \shortauthors{Finoguenov et al.}

\newcommand{\Rc}{\rm\,R_{C}}

\newcommand{\sqdegr}{\raisebox{0.65ex}{\tiny\fbox{$ $}}\,$^{\circ}$}

\begin{document}

\submitted{The Astrophysical Journal, 703:1-12, 2009}

\title{The roadmap for unification in galaxy group selection:. I. A search
  for extended X-ray emission in the CNOC2 survey.$^{*}$}

\author{A.~Finoguenov\altaffilmark{1,2}, J.~L.~Connelly\altaffilmark{1},
  L.~C.~Parker\altaffilmark{3}, D.~J.~Wilman\altaffilmark{1},
  J.~S.~Mulchaey\altaffilmark{4}, R.~P.~Saglia\altaffilmark{1}, 
  M.~L.~Balogh\altaffilmark{5}, R.~G.~Bower\altaffilmark{6},
  S.~L.~McGee\altaffilmark{5}}

\altaffiltext{$\star$}{Based on observations with the ESA/NASA {\it XMM-Newton}
  science mission; the European Southern Observatory, Chile; NASA/ESA
  {\it Chandra X-ray observatory} }

\altaffiltext{1}{Max-Planck-Institut f\"ur Extraterrestrische Physik,
             Giessenbachstra\ss e, 85748 Garching, Germany}
\altaffiltext{2}{University of Maryland, Baltimore County, 1000
  Hilltop Circle,  Baltimore, MD 21250, USA}
\altaffiltext{3}{Department of Physics \& Astronomy, McMaster University, 1280 Main St. W, Hamilton ON L8S 4M1, Canada}
\altaffiltext{4}{The Observatories of the Carnegie Institution of
  Washington, CA, USA}
\altaffiltext{5}{Department of Physics and Astronomy, University of Waterloo,
200 University Ave, Waterloo, ON N2L 3G1, Canada}
\altaffiltext{6}{Institute for Computational Cosmology, University of Durham, Durham, UK}


\begin{abstract}

  X-ray properties of galaxy groups can unlock some of the most challenging
  research topics in modern extragalactic astronomy: the growth of structure
  and its influence on galaxy formation. Only with the advent of the Chandra
  and XMM facilities have X-ray observations reached the depths required to
  address these questions in a satisfactory manner.  Here we present an
  X-ray imaging study of two patches from the CNOC2 spectroscopic galaxy
  survey using combined Chandra and XMM data.  A state of the art extended
  source finding algorithm has been applied, and the resultant source
  catalog, including redshifts from a spectroscopic follow-up program, is
  presented.  The total number of spectroscopically identified groups is 25
  spanning a redshift range 0.04--0.79. Approximately 50\% of CNOC2
  spectroscopically selected groups in the deeper X-ray (RA14h) field are
  likely X-ray detections, compared to 20\% in the shallower (RA21h) field.
  Statistical modeling shows that this is consistent with expectations,
  assuming an expected evolution of the L$_X$--M relation. A significant
  detection of a stacked shear signal for both spectroscopic and X-ray
  groups indicates that both samples contain real groups of about the
  expected mass.  We conclude that the current area and depth of X-ray and
  spectroscopic facilities provide a unique window of opportunity at
  $z\sim0.4$ to test the X-ray appearance of galaxy groups selected in
  various ways. There is at present no evidence that the correlation between
  X-ray luminosity and velocity dispersion evolves significantly with
  redshift, which implies that catalogs based on either method can be
  fairly compared and modeled.
\end{abstract}


\section{Introduction}

Groups of galaxies constitute the most common galaxy associations,
containing as much as 50-70\% of the galaxy population at the present day
(Geller \& Huchra 1983; Eke et al. 2005). Given that most galaxies will
encounter the group environment during their lifetime, an understanding of groups
is vital to understanding galaxy evolution in general.  The characteristic
depth of the potential wells of groups is similar to those of individual
galaxies, and the velocities of member galaxies are only a few hundred km/s.
Under these circumstances, galaxies interact strongly with one another and
also with the larger scale environment.  Recent simulation work suggests
that strangulation and ram-pressure stripping may quench star formation in
galaxies inhabiting groups with a significant intragroup medium component
(Kawata \& Mulchaey 2008; McCarthy et al. 2008).  Hence, not only are groups
the most common environment for galaxies, but also they provide an environment
which may have a strong effect on galaxy properties.

Perhaps the best way to understand the role the group environment plays in
galaxy evolution is to study the galaxy populations in groups over a range
of cosmic time. Historically, such studies have been limited because it has
been difficult to define group samples out to even moderate redshifts.  This
situation is rapidly changing, however.  The recent completion of very large
redshift surveys now allows large group samples to be kinematically-defined
out to z $\sim$ 1 (Carlberg et al. 2001; Gerke et al. 2005). The current
generation of X-ray telescopes also allows identification of X-ray groups
over a similar redshift interval (Willis et al. 2005; Finoguenov et al.
2007).

Follow-up studies of these group samples suggest that indeed there has been
considerable evolution in the group environment since z $\sim$ 1. In
spectroscopically-selected groups, there is evidence for a significant drop
in the fraction of strongly star-forming galaxies over the last few billion
years (Wilman et al. 2005b).  This process is further accompanied by an
apparent increase in the population of S0 galaxies (Wilman et al. 2009).
Recent studies show that, unlike their low redshift counterparts, many X-ray
luminous groups at intermediate redshift lack a dominant central galaxy
(Mulchaey et al. 2006; Jeltema et al. 2006, 2007). This implies that even
the most massive groups at intermediate redshifts are still in the process
of evolving.

Despite these first efforts to study groups over a range of cosmic time, a
coherent picture for the role they play in galaxy evolution has proved
elusive.  One of the reasons for this is that groups are a very
heterogeneous class of objects: they span a wide range of dynamical states
from objects just in the process of collapsing for the first time (like the
Local Group) to fully virialized systems with properties much like galaxy
clusters (e.g., Zabludoff \& Mulchaey 1998; Rasmussen et al. 2006).
Unfortunately, none of the aforementioned studies have included groups
across this complete range of dynamical states. This is largely because of
selection effects: groups selected to be X-ray luminous are biased toward
more massive, evolved and relaxed systems, while spectroscopically-selected
samples are dominated by small associations of a few galaxies that would
normally be undetectable in X-ray emission.  In addition, few studies have
had both good membership information (which requires extensive spectroscopic
programs) and X-ray data (the best indicator of a group's dynamical state)
and both of these are needed to obtain a comprehensive understanding of the
group environment.

Unfortunately, most groups are not X-ray luminous and therefore relatively
long X-ray observations are required to study these objects.  For this
reason, most studies have been restricted to groups that are a priori known
to be X-ray bright. Much less is known about the X-ray properties of the
more common systems that dominate spectroscopic group catalogs. In fact, the
fraction of spectroscopically-selected groups that contain a significant
X-ray emitting intragroup medium is still very poorly constrained (Mulchaey
2000). XMM-Newton studies of a few spectroscopically-selected groups suggest
their properties may be quite different from those of X-ray luminous systems
(Rasmussen et al. 2006).

An additional problem with studying groups is that it is difficult to
accurately determine the masses of these systems.  One promising technique
is to use weak lensing measurements, which has been extensively used as a
tool for measuring the masses and dark matter profiles of galaxy clusters
(e.g., Fahlman et al. 1994; Luppino \& Kaiser 1997; Hoekstra et al.  1998;
Clowe et al. 2006; Kubo et al. 2007) and the statistical studies of
galaxy-sized halos (Brainerd et al. 1996; Hudson et al. 1998; Hoekstra et
al. 2005; Mandelbaum et al. 2006b; Parker et al. 2007).  This technique can
also be employed for studies of galaxy groups (Hoekstra et al. 2001; Parker
et al. 2005; Mandelbaum et al. 2006b) though few studies have been carried
out, largely due to a lack of large samples of galaxy groups at intermediate
redshifts. Unlike rich galaxy clusters, individual galaxy groups do not
produce a measurable weak lensing signal and therefore the signal from
multiple groups must be stacked, much like the techniques employed in
galaxy-galaxy lensing.

Our approach to these problems is to perform a deep X-ray survey in an area
of the sky where an extensive spectroscopic survey, and deep optical
imaging, has been completed. Our X-ray observations have been designed to
guarantee the detection of X-ray groups down to low X-ray luminosities and
out to intermediate redshifts where the spectroscopic coverage is complete
enough to allow groups to be spectroscopically identified. By combining the
optical and X-ray data, we can select groups over the full range of
dynamical states.  Furthermore, by restricting our analysis to intermediate
redshifts, we can take advantage of the deep imaging to estimate weak
lensing masses for these groups.

The paper is structured as follows: in \S\ref{G} we recall the construction
of the sample of spectroscopic groups; in \S\ref{D} we describe the archival
and our own XMM-Newton and Chandra observations of the field and present the
details of the image processing technique and selection of extended X-ray
systems; an ongoing, dedicated spectroscopic follow-up program is outlined
in \S\ref{F} and the catalog of X-ray systems which could be unambiguously
identified is introduced in \S\ref{C}. In \S\ref{L} we describe the stacked
weak lensing analysis and provide the characteristics of various group
samples and in \S\ref{V} we discuss which spectroscopic groups are
identified in X-rays. We summarize our results in \S\ref{S}.  The goal of
this paper is to present the extended X-ray source catalog and provide a
template for comparison with the spectroscopic group catalog.  Results
presented here are part of an ongoing program: additional spectroscopic,
deeper X-ray and multiwavelength imaging data will provide improved
statistics and diagnostics for future papers.

All through this paper, we shall adopt a ``concordance'' cosmological model,
with $H_o=72$ km s$^{-1}$ Mpc$^{-1}$, $\Omega_M=0.25$, $\Omega_\Lambda =
0.75$, and --- unless specified --- quote all X-ray fluxes in the [0.5-2]
keV band and rest-frame luminosities in the [0.1-2.4] keV band, provide the
confidence intervals at the 68\% level and evaluate the enclosed density in
a definition for masses and radii in respect to the critical density.

\section{CNOC2 Survey and Groups}\label{G}

The second Canadian Network for Observational Cosmology Field Galaxy
Redshift Survey (CNOC2) is one of the few completed surveys with
spectroscopic and photometric data for a large, well-defined sample of
galaxies, obtained for the purpose of studying the evolution of galaxy
clustering (Yee et al.\ 2000). CNOC2 is magnitude limited down to
$\Rc\sim23.2$ in photometry and totals $\sim$1.5\sqdegr\ within four
separate patches on the sky. Spectroscopic redshifts, mainly in the range
0.1$\leq z \leq$0.55 exist for a large and unbiased sample of $\Rc \lesssim
21.5$ galaxies (Yee et al.\ 2000). Galaxy groups have been identified by
applying a friends of friends algorithm to detect significant overdensities
in redshift space, with parameters tuned to pick up virialized systems
(Carlberg et al. 2001): in practice this means that group selection is tuned
such that at least three galaxies with CNOC2 redshifts are clustered tightly
enough that they appear at least 200 times overdense with respect to the
critical density.  The resulting sample contains over 200 systems in
$\sim$1.5\sqdegr\ spanning a broad range of dynamical states.

To study the evolution of galaxies in the group environment, we undertook an
extensive follow-up program of a subset of these groups using the Magellan
6.5-m telescopes in Chile (Wilman et al. 2005a), HST, Spitzer, Chandra,
XMM-Newton and GALEX among other facilities. This extensive dataset allows
us to study the properties of the galaxies and the groups in detail. These
data provide ample evidence for significant changes in the star formation
rates and morphological composition of groups galaxies over the last 5
billion years, and differences from the field population (Wilman et al.
2005a,b, 2008, 2009; Balogh et al. 2007, 2009; McGee et al. 2008).

\section{X-ray data and methods for extended emission search}\label{D}

X-ray data have been obtained for two of the four CNOC2 patches described in
\S\ref{G} (the RA14h and RA21h patches).  Each of these patches has been
observed on several occasions with both XMM-Newton and Chandra. The
XMM-Newton OBSIDs for the RA14h patch are 0148520101, 0148520301, and
0149010201, while the Chandra OBSIDs for this patch are 5032, 5033 and 5034.
The XMM-Newton OBSIDs for the RA21h patch are 0404190101 and 0404190201 and
the Chandra OBSID for this patch is 6791.

To increase our sensitivity to low level X-ray emission, our analysis is
performed on X-ray mosaics made from the co-addition of the XMM-Newton and
Chandra data.  Throughout this paper, we list the effective exposure times
in each field in units of the equivalent Chandra exposure that would be
required to reach the observed sensitivity in the 0.5--2 keV band for a 2
keV thermal diffuse emission at z=0.2.  The maximum effective exposure times
in these units are 348~ksec in the RA21h patch and 469~ksec in the RA14h patch.
We note, that in calculating the properties of systems with a known
redshift, we self-consistently re-estimate the effective sensitivity,
accounting for the measured redshift and expected temperature (using $L_X$;
see Finoguenov et al. 2007). In Fig.1 we present the exposure map for both
patches. The primary differences between the X-ray coverage of the two
fields is that the depth in the RA14h patch is more uniform, while most of the
RA21h area has an effective exposure much less than the maximum (most of the
field has an effective exposure time of $\sim$ 100 ksec or less).

\begin{figure*}

\includegraphics[width=18.0cm]{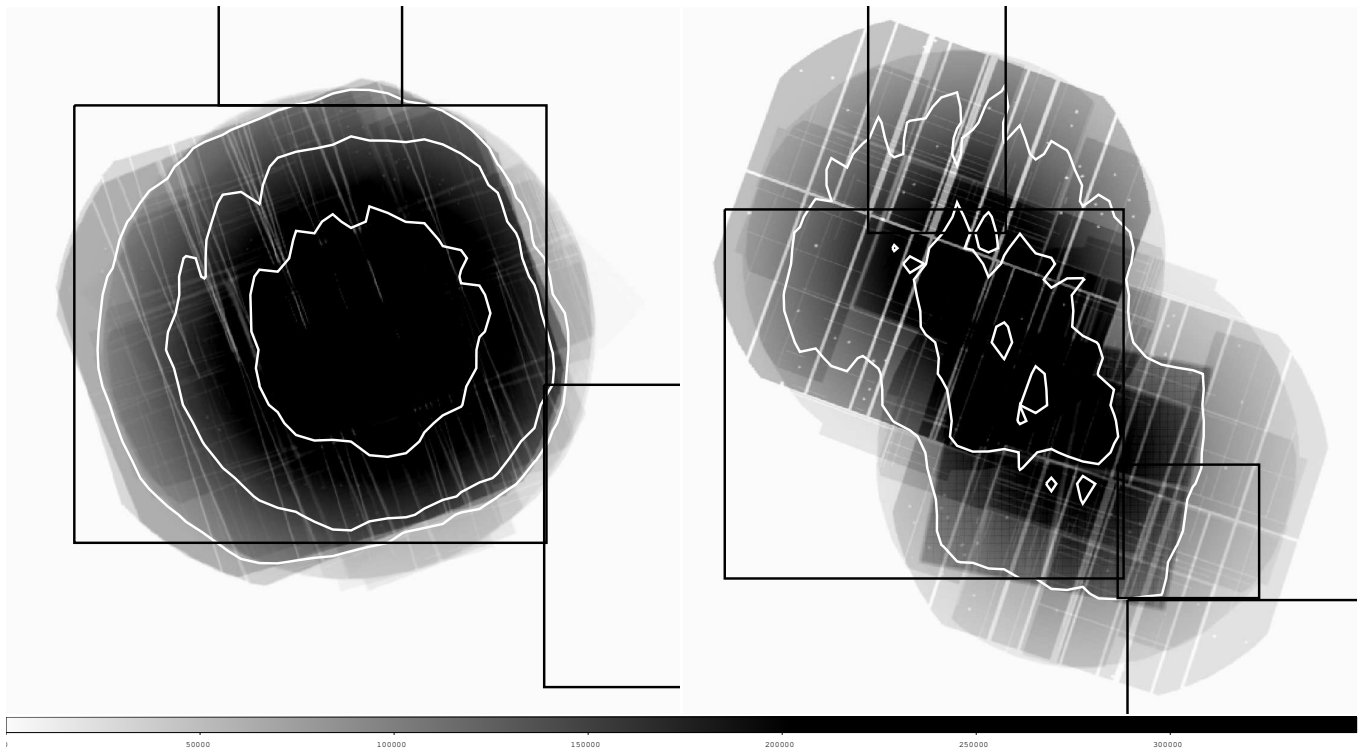}\hfill

\centering \figcaption{Exposure maps of RA14h (left) and RA21h (right) patch
  of CNOC2. Co-added XMM and Chandra exposures are shown. The contours
  (white) indicate the achieved level of effective 100, 200 and 300 ksec
  Chandra depth in the 0.5--2 keV band for a 2 keV source at z=0.2 (see
  text).  The total area covered by X-ray observations is 0.2 and 0.3 square
  degrees for the RA14h and RA21h fields, respectively. The black boxes show
  the boundary of the CNOC2 redshift survey.}
\end{figure*}

\subsection{XMM routine for point source removal}

In making the X-ray mosaic, we use a new procedure, which has now been
applied to several deep X-ray surveys (COSMOS, SXDF, LH, CDFS, CDFN, PISCES
fields). A direct comparison between XMM and Chandra catalogs, possible in
CDFS\&N, yields 80-90\% agreement in the source catalogs, with some residual
contamination due to alignments of AGNs that were unresolved in the XMM
dataset but were resolved with Chandra (Finoguenov et al. in prep.). The XMM
data reduction was done using XMMSAS version 6.5 (Watson et al.  2001;
Kirsch et al. 2004; Saxton et al. 2005). In addition to the standard
processing, we perform a more conservative removal of time intervals
affected by solar flares, following the procedure described in Zhang et al.
(2004) (see Finoguenov et al. 2007 for details).  In order to increase our
sensitivity to extended, low surface brightness features, in addition to
adopting the background subtraction procedure of Finoguenov et al.  (2007),
we also removed all hot MOS chips. From our careful monitoring of the
quality of the background subtraction, we confirm the findings of Snowden et
al. (2008) that episodically one chip in either MOS1 or MOS2 exhibits a
distinctly higher background below 1 keV.  This effect is particularly
problematic for recent XMM observations.

After the background has been estimated and subtracted for each observation
and each instrument separately, we produce the final mosaic of cleaned
images and correct it with a mosaic of the exposure maps. In this process,
we also account for the differences in sensitivity between the pn \& MOS
detectors.

As we are interested in identifying extended X-ray sources, adequate point
source removal is vital. The formal confusion limit for the XMM PSF
precludes using large scales for source detection at exposures exceeding 100
ksec.  However, since the shape of PSF is not Gaussian, the information on
large scales can be restored using the detection of point sources on small
scales and subsequent modeling of the point source flux. To illustrate the
point, since the observed image is a convolution of the original image and
the instrument PSF, we present the arguments in the Fourier space, where the
convolution is a mere multiplication of the Fourier transforms. As the
Fourier transform of a Gaussian is also a Gaussian, this very effectively
suppresses any information on scales below the width of the distribution.
The XMM PSF can be approximated by a sum of 3 Gaussians, so a convolution
with such a function is a sum of 3 convolutions, proving that the
information is still retained on scales larger than the width of the core of
PSF ($5^{\prime\prime}$). This defines the theoretical lower limit on
recovering the information content in XMM images.

It is also obvious that the removal of point source flux has to be done
prior to examining the confused spatial scales. To accommodate this, we
introduced such a {\em new} process to identify the point sources and to
remove their flux. The procedure adopts a symmetric model for the XMM PSF
and uses the calibrations of Ghizzardi \& Molendi (2002, see also Ferrando
et al. 2003).  There are three steps of the flux removal, which are
wavelet-specific, and their goal is to achieve a correct flux estimate for
each detected point source.  The wavelet decomposition we employ is both a
spatial and a significance filter, so flux is stored and subtracted from the
input image, before applying the next spatial scale filtering, only if its
significance is larger than some threshold (Vikhlinin et al. 1998). As a
result, in the wavelet decomposition the flux attributed to each spatial
scale varies: as the significance of the source increases, more flux is
attributed to smaller scales. A simplified procedure would result in either
underestimation of flux pollution from the faint sources or in the
overestimation of the pollution from the strong sources.  To properly
subtract the PSF-induced contribution of small-scales to large-scale
features, we perform three reconstructions of the small scales, using the
detection thresholds of 100, 30 and 4 $\sigma$. The $4\sigma$ image receives
the full weight according to PSF models and the other two reconstructions
are subtracted from it with proper weights, which are defined by both the
XMM PSF on small scales and properties of the wavelet transformation. A
similar and even better result can be obtained by fitting the PSF to the
positions of the sources, which we hope to include in future analysis.  The
XMM PSF model describing the detections on small scales also predicts the
flux on large scales. As discussed in Finoguenov et al.  (2007), our
selection of spatial scales is done to reduce the variation of this
prediction with off-axis angle and greatly simplify the procedure of point
source flux removal.  To subtract the expected flux spread from small to
large scales, we convolved the point source image reconstructed above with
Gaussians of width 16, 32 and 64 arcseconds, weighted according to the
observed PSF and then subtracted this from the image together with the
wavelet reconstruction on the small scales.  To account for deviations
between the symmetric PSF model and a two-dimensional PSF characterization,
we add 5\% systematic error associated with our model on the 16 and 32
arcsecond scales and 200\% systematic error associated with our model on the
64 arcsecond scale. The systematic errors correspond to differences
occurring at the edge of the field of view.  Using XMMSAS detections of
point sources in the Lockman Hole, the model XMM images of point sources in
that field have been reconstructed (Brunner et al.  2008).  We run our
procedure on these randomized images detecting no residual extended
emission.

\begin{figure*}

\includegraphics[width=18.cm,clip]{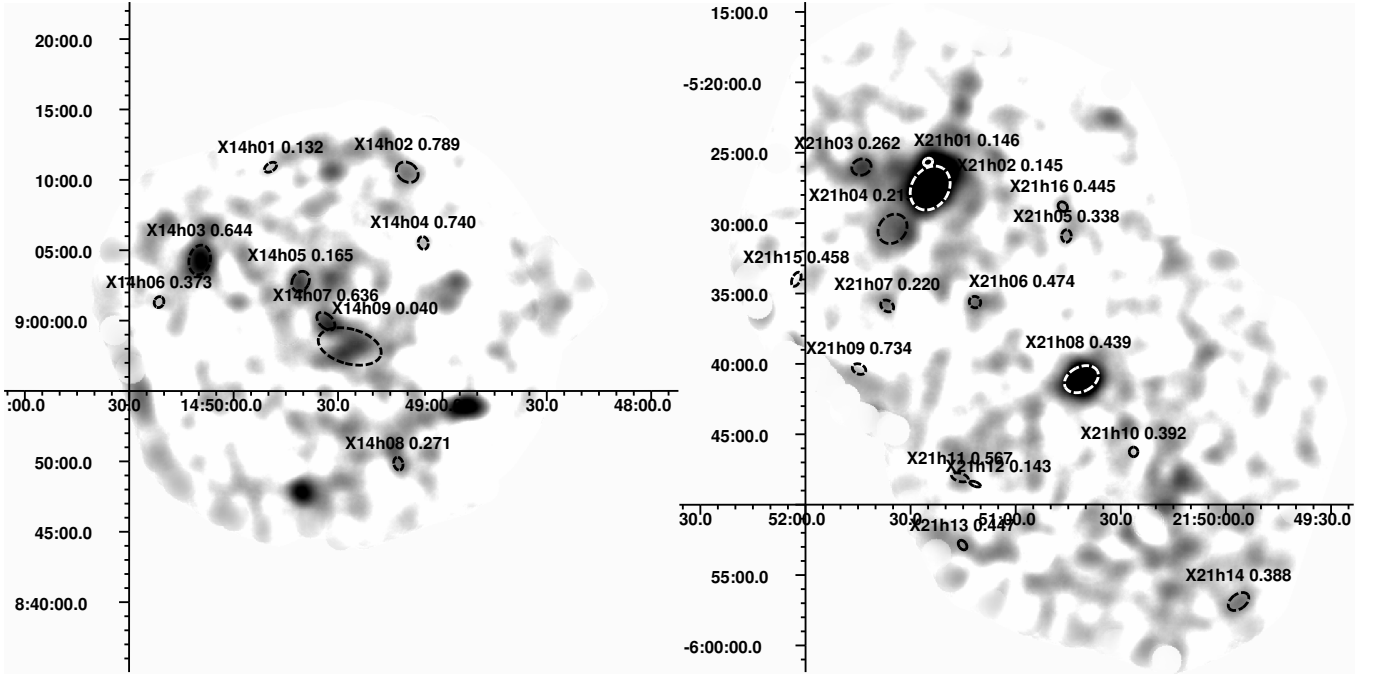}

\centering \figcaption{Signal-to-noise ratio of the point-source removed
  X-ray images in the 0.5--2 keV energy range of the RA14h (left) and the RA21h
  (right) CNOC2 patches. The gray scale indicates the significance per
  $32^{\prime\prime}$ beam starting at 1 $\sigma$ (light gray).  Black
    color indicates a significance in excess of 7 sigma. Ellipses
  indicate the position and the size of the flux extraction regions used to
  estimate the properties of identified systems. In addition to the sources
  detected in the final wavelet reconstruction, 6 low significance X-ray
  sources from Tab.1 are shown for completeness. Some of the
  high-significance X-ray peaks are still to be spectroscopically
  identified. }
\end{figure*}

\begin{figure*}
\includegraphics[width=18.cm,clip]{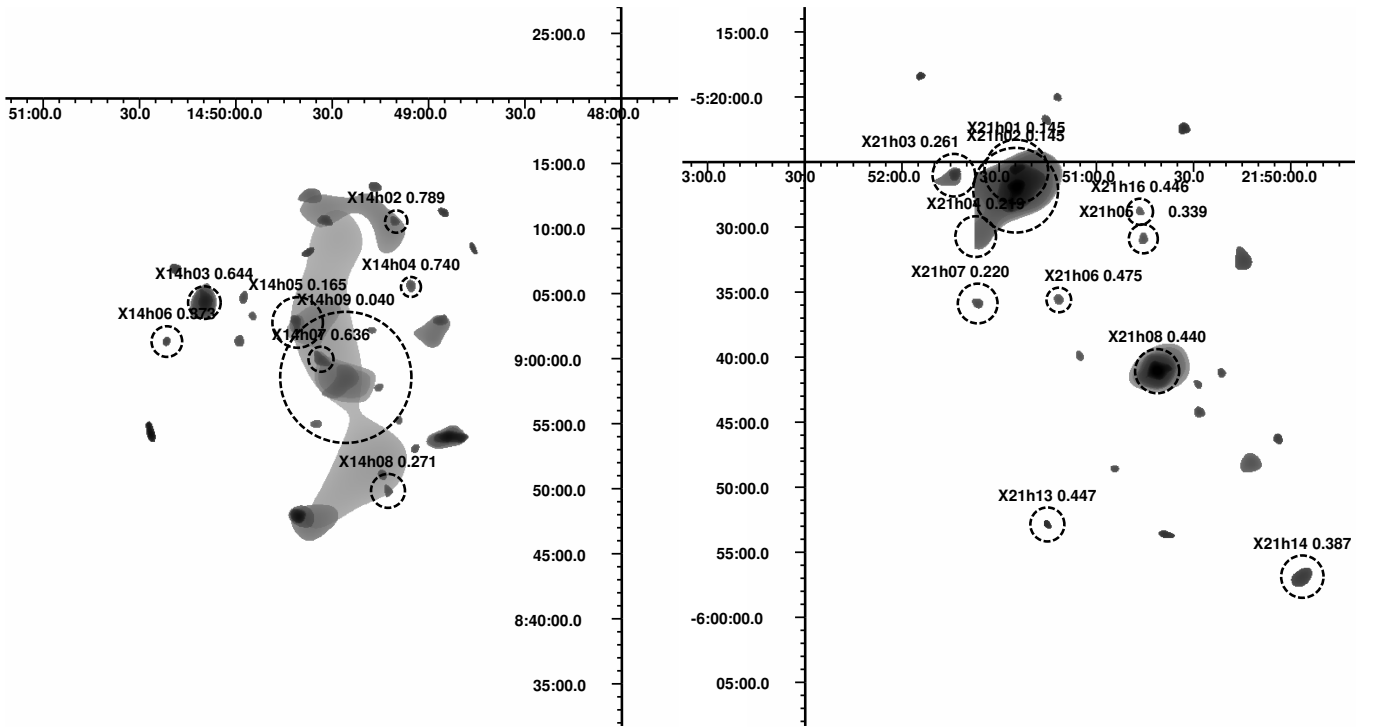}
\centering \figcaption{Wavelet-reconstruction of the point-source removed
  X-ray image of the RA14h (left) and the RA21h (right) CNOC2 patches on
  spatial scales from 32$^{\prime\prime}$ to 256$^{\prime\prime}$. The units
  of the image are ergs s$^{-1}$ cm$^{-2}$ arcmin$^{-2}$ and the logarithmic
  range shown is from -16 to -13 (from light gray to black). Circles
  indicate the spectroscopically identified X-ray systems with radii of the
  circle showing the $R_{500} (\approx 0.6R_{200}$). In most cases the
  emission covers only the core of the detected systems.  Some of the
  high-significance X-ray peaks are still to be spectroscopically
  identified.}
\end{figure*}

\subsection{Chandra data reduction}

Initial Chandra data reduction was performed using standard reduction
procedures of CIAO version 3.4\footnote{http://cxc.harvard.edu/ciao/}.  In
this work we use the available Chandra observations of the CNOC2 fields to
further improve the sensitivity toward the detection of extended X-ray
emission. Since we use the $32^{\prime\prime}$ and $64^{\prime\prime}$
scales to search for extended emission with XMM, we similarly removed the
emission on smaller spatial scales from Chandra data. Large scales
considered for extended source detection allowed us to use the full Chandra
field for this work. We removed point sources from the Chandra data using
the following procedure.  For Chandra, the on-axis scattering of point
source flux into $32^{\prime\prime}$ scales is negligible, while it becomes
important at off-axis angles exceeding $3^\prime$. Our simplified point
source removal is based on the PSF model which shows that once the scales of
$8^{\prime\prime}$ and $16^{\prime\prime}$ are polluted, so are the larger
scales. Residual variations in the off-axis behavior of the Chandra PSF are
treated as systematic errors in our model. In summary, in addition to the
removal of point source flux detected on scales of 1, 2, 4, 8 and 16
arcseconds, we use the emission detected at scales of $8^{\prime\prime}$ and
$16^{\prime\prime}$ to predict and subtract the effect of the Chandra PSF,
important for off-axis angles exceeding $3^\prime$.  We added in quadrature
a 20\% systematic error associated with this model to the error budget.

\subsection{Combined X-ray imaging}

After instrument--specific background and point source removal, the residual
images were co-added, taking into account the difference in the sensitivity
of each instrument to produce a joint exposure map.  Specifically, the
Chandra ACIS-I exposure is taken as it is, each XMM EPIC MOS exposure is
counted as equal to Chandra exposure, and the XMM EPIC pn exposure is
multiplied by 3.6 times the read-out time correction (0.93--0.98 depending
on the read-out mode).

The resulting signal-to-noise images are shown in Fig.2, together with the
location of identified systems and the flux extraction regions.  To detect
the sources we run a wavelet detection at $32^{\prime\prime}$ and
$64^{\prime\prime}$ spatial scales, similar to the procedure outlined in
Finoguenov et al. (2007). The total number of detections in the RA14h and
RA21h patches is 27 and 23, respectively.  This corresponds to a source
surface density of 135 and 75 per square degree in the two fields,
respectively.

The refined procedure of point source flux removal has allowed us to better
associate the peak of the X-ray emission with the center of the group.  The
formal positional uncertainty is of order of $10^{\prime\prime}$, although
it can reach $30^{\prime\prime}$ for systems of low statistical significance
due to X-ray flux contribution from sources under the detection threshold.
The sources flux is measured from the residual image after the background
and point sources have been subtracted off.  This means that the estimate of
flux signal to noise ratio is not the same as the significance of the source
detection (estimated using the wavelet image).

\section{Additional Optical Spectroscopy}\label{F} 

The original CNOC2 survey (Yee et al. 2000) has only a $\sim30$\% sampling
rate, as well as a significant spatial variation in spectroscopic
completeness.  Wilman et al.  (2005a) improved the completeness in the
regions around a subset of spectroscopically defined groups from Carlberg et
al.  (2001).  However, by definition this was in regions where the
completeness was already high enough to find the group in redshift space.
Many of the X-ray detected systems therefore exist in regions of highly
incomplete spectroscopy, such that it is impossible to determine the group
redshift from the existing spectroscopy.  Therefore, to determine the
redshifts of the X-ray detected systems, we have obtained additional
spectroscopy with the VLT and Magellan telescopes. This substantially
improves the sampling rate in regions of X-ray emission, increasing the
number of group identifications and known membership. The spectroscopic data
and group membership will be described in full by Connelly et al. (in
prep.), but a brief discussion of new redshift measurements is provided
here.

The VLT observations were conducted with FORS2 over the course of three
visitor mode observing runs in 2007-2008, with corresponding run IDs of
080.A-0427(D) (0.6 night, Oct 5 2007), 080.A-0427(B) (two half nights
starting Mar 1, 2008) and 081.A-0103(B) (two half nights starting Aug 24,
2008).  A total of 21 MXU (multi-object) masks
(6.8\arcmin~$\times$~6.8\arcmin\ FOV) were observed.  Observations were
obtained in both the RA14h and RA21h fields, and were designed to maximize
the number of extended X-ray sources targeted.  Slits were placed on
galaxies with unknown redshifts, prioritizing galaxies close to the X-ray
centers and with magnitudes $R_C \lesssim 22.0$, although fainter and more
distant galaxies were used to fill the masks.  The spectral set-up of the
observation used the GRIS300V grism and GG375 filter resulting in an
effective wavelength range of $\sim430-700$ nm. A slit width of
$1^{\prime\prime}$ was used for all objects.  The total integration time per
mask was approximately 1 hr.  Chip images from consecutive mask exposures
were co-added using the IRAF {\it imcombine} tool with cosmic-ray rejection
applied.  The data reduction was performed with the most recent version of
the standard FORS pipeline which performs bias correction, flat-fielding,
correction for optical distortions, and wavelength calibration (Appenzeller
et al. 1998). The pipeline also detects and extracts individual object
spectra. Finally, galaxy redshifts were derived manually using the emission
lines where available, or the H and K Calcium lines and the G-band feature
for absorption redshifts. To date, we have measured redshifts for 364 (312
with high quality) objects from the FORS2 data. This focused on galaxies in
the cores of X-ray systems, measuring enough redshifts to be confident of
the group redshift.  Data reduction of a complete FORS2 dataset is ongoing
and further results will be presented by Connelly et al.  (in prep).

We also obtained two multi-object masks of the RA14h field using the IMACS
instrument on the Baade/Magellan I telescope in July 17-18, 2007. The IMACS
observations were taken with a grism of 200 lines mm$^{-1}$, giving a
wavelength range of $\sim$ 5000--9500\AA \ and a dispersion of 2.0 \AA \
pixel$^{-1}$. A slit width of $1^{\prime\prime}$ was used.  The exposure
time was 2 hour for both masks.  The IMACS data were reduced using the
COSMOS data reduction package. First, overscan regions of the CCDs were used
to measure and subtract the bias level. Domeflat exposures taken during the
night were used to flat-field the data. Sky subtraction was performed using
the method outlined in Kelson (2003). Wavelength calibrations were
determined from HeNeAr arc exposures.  Redshifts were measured by
cross-correlating the IMACS spectra with SDSS galaxy templates.  We measured
a total of 57 new redshifts from the IMACS data.

\section{A catalog of identified X-ray groups}\label{C}

\begin{deluxetable}{lcccccrrccccc}
\tablewidth{0pt}
\tabletypesize{\footnotesize}
\tablecaption{Current status of the spectroscopic follow-up of X-ray extended sources.}\label{t:ol}
\tablehead{
\colhead{ID } &
\colhead{IAU name} &
\colhead{R.A} &  
\colhead{Decl.} & 
\colhead{z} &
\colhead{$n_z$} &
\colhead{flux  $10^{-14}$} &
\colhead{L$_{\rm 0.1-2.4 keV}$} & 
\colhead{M$_{200}$} & 
\colhead{$R_{200}$} \\
\colhead{} & 
\colhead{CNOC2XGG J} & 
\multicolumn{2}{c}{Eq.2000} & 
\colhead{ } & 
\colhead{ } & 
\colhead{ergs cm$^{-2}$ s$^{-1}$} & 
\colhead{$10^{42}$ ergs s$^{-1}$} & 
\colhead{$10^{13}$ M$_\odot$} & 
\colhead{$\prime$} &
\colhead{optical ID}
}
\startdata
X14h01$^a$&144949+0910.9 & 222.45584 & +9.18167 & $0.132\pm0.002$ &5 & 0.95$\pm$0.31 & 0.66$\pm$0.21 & 1.68$\pm$0.33 & 3.76& \\
X14h02&144910+0910.5 & 222.29202 & +9.17579 & $0.789\pm0.003$ &4 & 0.27$\pm$0.07 & 14.31$\pm$3.56 & 6.94$\pm$1.06 & 1.85& \\
X14h03&145009+0904.3 & 222.54075 & +9.07188 & $0.644\pm0.001$ &9 & 1.24$\pm$0.13 & 36.15$\pm$3.83 & 14.32$\pm$0.95 & 2.57& \\
X14h04&144905+0905.5 & 222.27293 & +9.09198 & $0.740\pm0.003$ &4 & 0.14$\pm$0.05 & 6.42$\pm$2.39 & 4.35$\pm$0.98 & 1.63& \\
X14h05&144940+0902.7 & 222.41983 & +9.04645 & $0.165\pm0.001$ &6 & 0.89$\pm$0.15 & 1.01$\pm$0.17 & 2.16$\pm$0.23 & 3.38 & 1\\
X14h06&145021+0901.3 & 222.58939 & +9.02196 & $0.373\pm0.002$ &6 & 0.39$\pm$0.13 & 2.93$\pm$0.98 & 3.63$\pm$0.73 & 2.20 & 28\\
X14h07&144933+0859.9 & 222.38956 & +8.99939 & $0.636\pm0.004$ &3 & 0.32$\pm$0.05 & 10.02$\pm$1.89 & 6.34$\pm$0.74 & 1.97& \\
X14h08&144912+0849.8 & 222.30276 & +8.83071 & $0.271\pm0.001$ &10 & 0.35$\pm$0.10 & 1.26$\pm$0.36 & 2.30$\pm$0.40 & 2.35 & 11\\
X14h09&144925+0858.5 & 222.35742 & +8.97616 & $0.040\pm0.001$ &5 & 2.68$\pm$0.35 & 0.16$\pm$0.02 & 0.73$\pm$0.06 & 8.49 & \\
X21h01&215124$-$0525.6 & 327.85410 & $-$5.42783 & $0.146\pm0.001$ &11 & 2.10$\pm$0.38 & 1.72$\pm$0.31 & 3.08$\pm$0.34 & 4.25 & \\
X21h02&215124$-$0527.5 & 327.85155 & $-$5.45853 & $0.145\pm0.001$ &19 & 7.63$\pm$0.29 & 6.50$\pm$0.25 & 7.21$\pm$0.17 & 5.64 & 104\\
X21h03$^b$&215143$-$0526.0 & 327.93324 & $-$5.43371 & $0.262\pm0.001$ &8 & 0.95$\pm$0.18 & 3.14$\pm$0.58 & 4.15$\pm$0.48 & 2.94& \\
X21h04&215135$-$0530.4 & 327.89632 & $-$5.50686 & $0.219\pm0.001$ &12 & 0.57$\pm$0.12 & 1.21$\pm$0.26 & 2.34$\pm$0.31 & 2.77 & 117\\
X21h05&215045$-$0530.9 & 327.68966 & $-$5.51535 & $0.338\pm0.002$ &6 & 0.25$\pm$0.08 & 1.52$\pm$0.48 & 2.45$\pm$0.47 & 2.05& \\
X21h06&215111$-$0535.6 & 327.79834 & $-$5.59362 & $0.474\pm0.002$ &6 & 0.20$\pm$0.06 & 2.83$\pm$0.87 & 3.25$\pm$0.61 & 1.83& \\
X21h07&215136$-$0535.8 & 327.90274 & $-$5.59811 & $0.220\pm0.001$ &12 & 0.51$\pm$0.14 & 1.10$\pm$0.31 & 2.20$\pm$0.37 & 2.71& \\
X21h08&215041$-$0541.0 & 327.67186 & $-$5.68486 & $0.439\pm0.001$ &30 & 2.82$\pm$0.13 & 31.63$\pm$1.50 & 15.72$\pm$0.47 & 3.23 & 138\\
X21h09$^a$&215144$-$0540.3 & 327.93620 & $-$5.67284 & $0.734\pm0.004$ & 3& 0.17$\pm$0.08 & 7.81$\pm$3.66 & 4.96$\pm$1.38 & 1.70& \\
X21h10$^a$&215026$-$0546.2 & 327.60980 & $-$5.77096 & $0.392\pm0.001$ &11& 0.22$\pm$0.10 & 1.86$\pm$0.81 & 2.67$\pm$0.70 & 1.93& \\
X21h11$^a$&215115$-$0548.0 & 327.81629 & $-$5.80110 & $0.567\pm0.001$ &9& 0.18$\pm$0.09 & 4.03$\pm$2.04 & 3.76$\pm$1.16 & 1.75& \\
X21h12$^a$&215111$-$0548.5 & 327.79831 & $-$5.80904 & $0.143\pm0.002$ &5& 0.13$\pm$0.29 & 0.12$\pm$0.28 & 0.57$\pm$0.64 &2.45& \\
X21h13&215115$-$0552.8 & 327.81317 & $-$5.88123 & $0.447\pm0.002$ &6 & 0.82$\pm$0.29 & 10.01$\pm$3.56 & 7.48$\pm$1.62 & 2.50& \\
X21h14&214956$-$0556.8 & 327.48498 & $-$5.94811 & $0.388\pm0.002$ &4 & 1.86$\pm$0.46 & 15.46$\pm$3.79 & 10.40$\pm$1.57 & 3.05& \\
X21h15$^a$&215202$-$0533.9 & 328.01069 & $-$5.56647 & $0.458\pm0.003$ &4 & 0.18$\pm$0.16 & 2.41$\pm$2.16 & 2.98$\pm$1.51 & 1.81& \\
X21h16&215046$-$0528.8 & 327.69397 & $-$5.48061 & $0.445\pm0.001$ &12 & 0.23$\pm$0.07 & 2.89$\pm$0.85 & 3.38$\pm$0.61 & 1.92& \\
\enddata
\\
\end{deluxetable}

$^a$ -- not in the final X-ray catalog, not included in statistical tests.\\
$^b$ -- outside CNOC2 spectroscopic survey

X-ray group redshifts are identified where at least three consistent galaxy
redshifts (CNOC2 $+$ preliminary FORS2 and IMACS catalog) lie within the
observed extent of X-ray emission ($\sim $0.3$\times R_{200}$).  In many
cases, the clustering of galaxies in both redshift and projected spatial
coordinates is obvious.  In some cases a brightest group galaxy (BGG)
clearly exists near the X-ray center, but this is not required for group
identification. In Tab.~1, we list the 25 X-ray sources associated with
spectroscopically confirmed groups.  This list includes 6 groups, which were
detected at an earlier stage of detection routine development. The flux from
these sources is highly uncertain because of systematic uncertainties in the
PSF subtraction.  We also checked that, except for the X21h09 group, the
other 5 groups are still detected with the adopted version of the detection
routine, when the threshold for source detection is lowered to 2 sigma.
These sources cannot be used in statistical tests, but since their
identification has been successful, they provide an example of faintest
X-ray groups and the data in Tab.~1 can be used to access their parameters.
The final wavelet reconstruction of the point-source removed X-ray image in
the 0.5--2 keV band and the 19 spectroscopically identified
high-significance peaks are shown in Fig.3.

Total flux in the 0.5--2 keV band is estimated by extrapolating the surface
brightness to $R_{500}\sim0.6R_{200}$ following the model prescription of
Finoguenov et al. (2007).  Rest-frame luminosity in the 0.1--2.4 keV band is
estimated (Finoguenov et al. 2007; Leauthaud et al. 2009) and the total
group mass within the estimated $R_{200}$, $M_{200}$, is computed following
the $z\sim 0.25$ relation from Rykoff et al. (2008) and assuming standard
evolution of scaling relations: $M_{200} E_z = f(L_X E_z^{-1})$, where \mbox{$E_z=(\Omega_M (1+z)^3 +
\Omega_\Lambda)^{1/2}$}. The assumed
scaling relations for systems of similar mass and redshift have been
verified using a weak lensing calibration of X-ray groups in the COSMOS
survey (Leauthaud et al. 2009).

The total number of galaxies within $R_{200}$ and $|\Delta z| < 0.007(1+z)$
is computed, $n_z$. Such selection is relatively loose and will include
interlopers, and is done as a preliminary step before applying sigma
clipping. Group membership allocation will improve as more redshifts are
measured, also providing a better estimate of velocity dispersion and
rejection of outliers.  At this preliminary stage the measurement of $n_z$
reinforces the existence of overdensities at the position and redshift of
X-ray groups.  Tab.~1 provides cluster identification number (column 1); IAU
name (2); R.A.  and Decl.  of a global center of the extended X-ray emission
for Equinox J2000.0 (3 \& 4); spectroscopic redshift (5); the number of
member galaxies inside $R_{200}$, $n_z$ before any sigma-clipping (6); the
total flux in the 0.5--2 keV band (7); rest-frame luminosity in the 0.1--2.4
keV band (8); estimated group total mass, $M_{200}$ (9) and the
corresponding $R_{200}$ (10).  Uncertainties are quoted at the 68\%
confidence level and do not include the scatter in scaling relations.  The
final column (11) lists the group number from the spectroscopically selected
group catalog of Carlberg et al. where there is a confident match with the
X-ray detected system.  The X-ray system X14h09 is identified with a bright
early-type galaxy at a $z=0.04$: other than spatial coincidence, the very
large X-ray extent argues in favor of a low redshift source.  Four satellite
galaxies within the estimated $R_{200}$ support this hypothesis.  However it
is likely that emission from spectroscopic group 38 at $z=0.511$ is confused
with this foreground system (see \S\ref{V}).

\section{Lensing}\label{L}

A weak lensing analysis of a large sample of CNOC2 spectroscopically
selected groups has been carried out based on deep CFHT and KPNO 4-m data as
described in Parker et al. (2005). High quality R- and I-band data were used
to measure the shapes of faint background galaxies. The analysis for each of
the 4 CNOC2 fields was based on single-band photometry, so the redshifts for
the background sources had to be estimated based on the N(z) distribution
from the Hubble Deep Field (Fern{\'a}ndez-Soto et al. 1999).  Without
accurate redshifts for the background sources there could be some
contamination in the source catalogs from faint group members. However, we
do not find an excess source density around the groups, compared with the
field, which suggests that the level of this contamination is small relative
to other uncertainties.

In order to compare the weak lensing properties of the spectroscopically and
X-ray selected groups the tangential shear signal was recomputed for
spectroscopically selected groups within the 14 and 21 hour CNOC2 fields, as
well as the area within those two fields overlapping with the region
observed at X-rays. The source catalogs used in this analysis are identical
to those used in Parker et al. (2005) and have been thoroughly tested for
systematics. The stacked weak lensing results are presented in Fig.4a and
Tab.2. The tangential shear for a sample of group lenses can be used to
calculate the ensemble-averaged velocity dispersion, assuming an isothermal
sphere density profile, as follows

\begin{equation}
\gamma_T=\frac{\theta_E}{2\theta}=\frac{2\pi\sigma^2}{c^2 \theta }\frac{D_{LS}}{D_S}\label{eqn:gt}
\end{equation}

where $\theta_E$ is the Einstein radius of the lenses, $\theta$ is the
angular distance from the group center, $D_{LS}$ is the angular diameter
distance between the lenses and sources, and $D_S$ is the angular diameter
distance to the sources. The best fitting isothermal sphere yields a
velocity dispersion of 228$\pm$137 km s$^{-1}$ for all spectroscopically
selected groups in the two patches and 260$\pm110$ km s$^{-1}$ for those
within the region of the survey covered by X-ray data.  The sample of all
X-ray groups yielded a best fitting isothermal sphere with a velocity
dispersion of 247$\pm$138 km s$^{-1}$. The errors in velocity dispersion are
calculated from weighted fits to an isothermal sphere profile, where the
weights are defined by the errors in the shear measurements. The errors in
the shear estimates are determined from the uncertainties in the source
shape measurements as described in Hoekstra et al. (2000). The tangential
shear was also computed for two samples of X-ray selected groups: the groups
with spectroscopic redshifts described above, and the entire sample of
extended X-ray sources. These results are presented in Fig.4b and Tab.2.

The X-ray groups with identified spectroscopic redshifts have an
ensemble-averaged velocity dispersion of 309$\pm106$ km s$^{-1}$, 
consistent with the result for spectroscopically selected groups. 
We also stack all X-ray groups (including those with no confirmed 
redshift) which requires an assumption of their N(z) which we take 
from the modeling presented in Fig.6.

It is clear from Fig.4 that the isothermal sphere model does not do a
particularly good job of recovering the shear profile, and thus the formal
measurement of velocity dispersion and its associated error do not provide a
full description of the data.  For example, the low shear signal in the
inner 24\arcsec\ bin can be attributed to the uncertain centering of groups,
whereas beyond $\sim$120\arcsec\ the errors are larger than the expected
signal.  To rule out the null hypothesis of no shear signal we also compute
the weighted mean shear within a 120\arcsec\ fixed aperture and associated
error (Table~2), where the weights come from the errors in the individual
shape measurements.  It is apparent that the no mass hypothesis can be ruled
out at the $\gtrsim3\sigma$ level for spectroscopically selected groups and
spectroscopically identified X-ray groups, while the uncertainty introduced
by assuming N(z) for all X-ray systems is greater than the increased
precision by having more groups. The total shear and the model based
ensemble-averaged velocity dispersion of identified systems is consistent
with that of spectroscopically selected groups within the errors.

The stacked signal at the position of X-ray extended sources is completely
dominated by the RA21h field groups ($3\sigma$).  There is no clear weak
lensing detection in the RA14h field alone ($<1\sigma$).  We have tested the
influence of the most massive $10^{14}M_{\odot}$ ($7\times10^{13}M_{\odot}$)
systems in the RA21h field by removing them from the sample and recomputing
the shear. The measured shear within 120\arcsec\ is still positive at the
$3\sigma$ ($2\sigma$) level, suggesting the real detection of lower mass
systems. However we note that in the RA14h field only 3 identified systems of
high enough mass for X-ray detection exist at $0.2<z<0.5$, the optimal range
for lenses at the depth of our data.  This suggests a role for cosmic
variance driving the difference between the two fields.  We have also
attempted a detection of stacked shear signal from the 13 spectroscopic
groups located inside the X-ray survey for which we are confident that there
is no X-ray emission.  The obtained result is listed in Tab.2.  While the
isothermal model fit is easily consistent with zero, there is a
$\sim2\sigma$ significant detection of shear within 120\arcsec.  This can
tentatively be taken as detection of non X-ray bright groups, although we
stress the low number statistics and suggest that much better statistics
would be required for confident weak lensing detection of mass in groups
beyond our X-ray detection limit.

\begin{table*}[htdp]
\caption{Weak lensing results}
\begin{center}
\begin{tabular}{c c c c c}
\hline
\hline
Sample & $N_{groups}$ & Mean z & Mean shear ($<$2\arcmin) & $\sigma$ km s$^{-1}$  \\
\hline
Spectroscopic groups in RA21h \& RA14h fields & 70 & 0.33 & 0.077 $\pm$ 0.024 & 228 $\pm$ 137 \\
Spectroscopic groups within X-ray survey & 35 & 0.33 & 0.099 $\pm$ 0.028 & 260 $\pm$ 110 \\
Spectroscopic groups with no X-rays & 13 & 0.33 & 0.100 $\pm$ 0.045 & 101 $\pm$ 326 \\
All X-ray detected systems & 49 & N(z) & 0.032 $\pm$ 0.025 & 247 $\pm$ 138 \\
X-ray groups with redshifts & 24 & 0.41 & 0.090 $\pm$ 0.035 & 309 $\pm$ 106 \\
\hline
\end{tabular}
\end{center}
\label{default}
\end{table*}%

\begin{figure*}
\includegraphics[width=9.0cm]{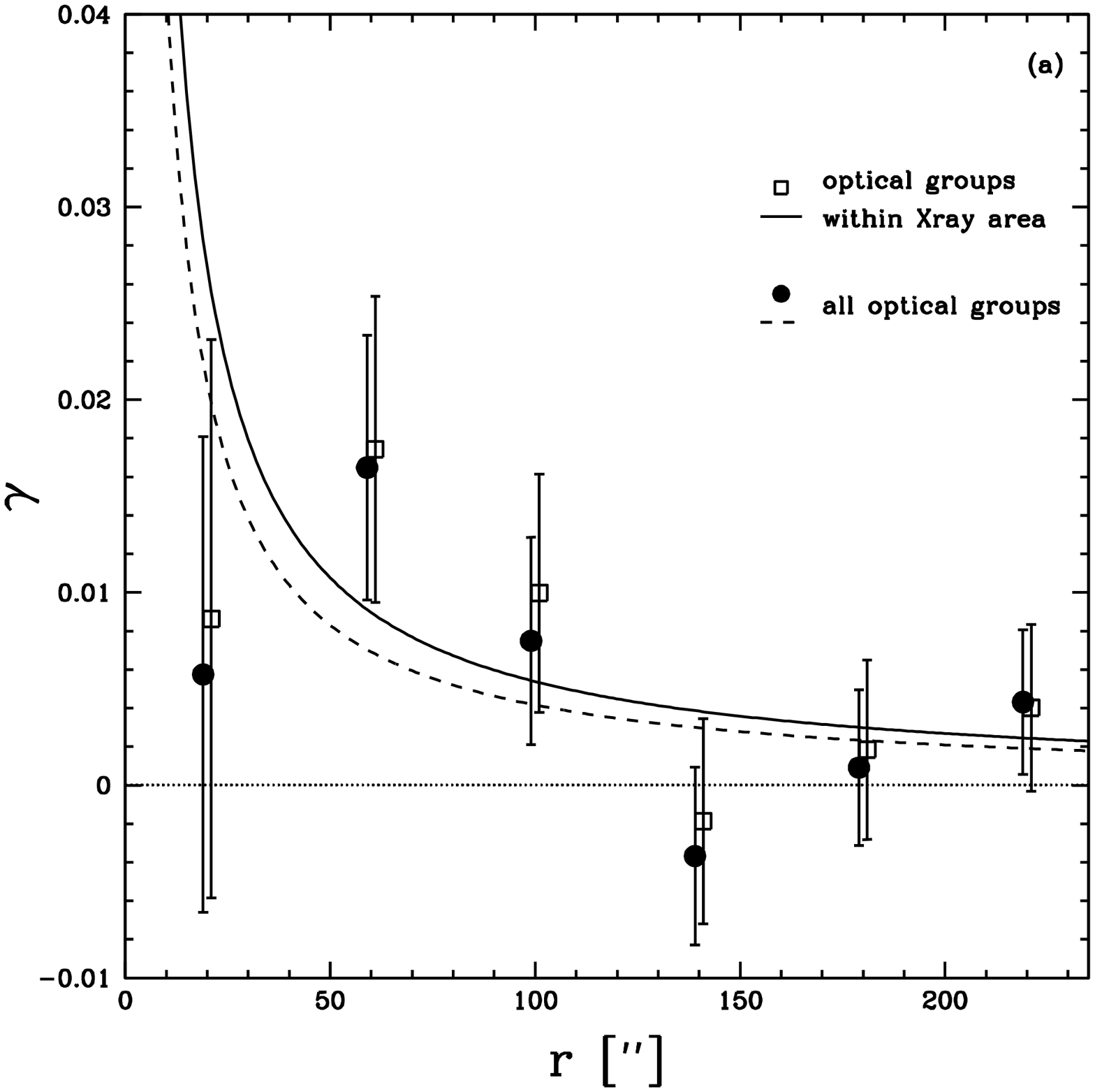}\hfill
\includegraphics[width=9.0cm]{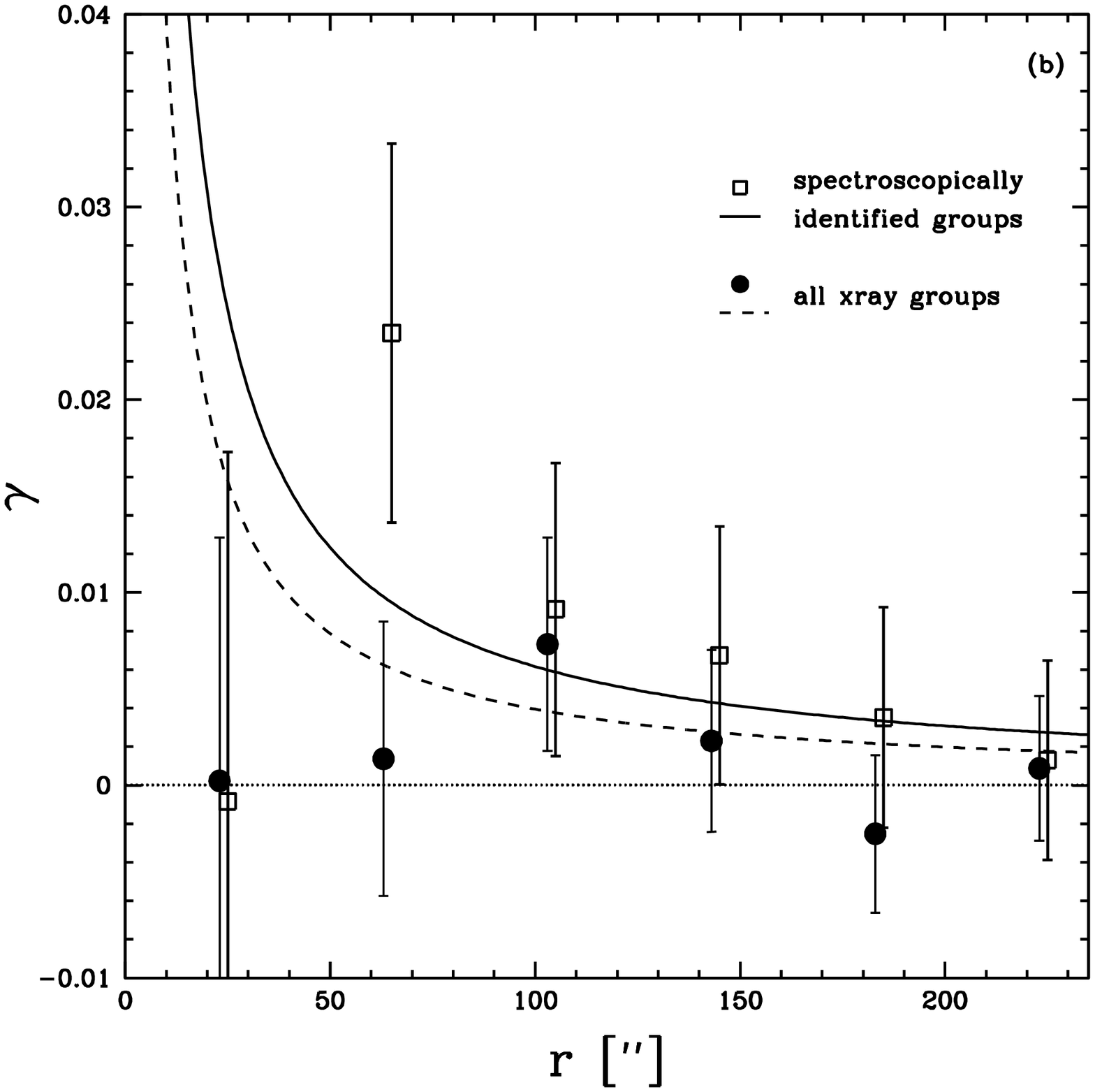}
\centering \figcaption{a) Stacked tangential shear profile of
  spectroscopically selected and b) X-ray selected groups. Overplotted lines
  are the best fits for an isothermal sphere model.}
\end{figure*}

\section{Comparison between spectroscopic and X-ray groups}\label{V}

Historically, X-ray surveys of spectroscopically selected group samples have
been limited by the sensitivity of X-ray telescopes. For this reason, very
few spectroscopic groups have known X-ray counterparts.  As we will demonstrate below, X-ray
observations have achieved the depths required to study $10^{13} M_\odot$
groups at intermediate redshifts. At the same time, galaxy redshift
surveys have also improved yielding much cleaner group catalogs and better
overall agreement between spectroscopically and X-ray selected group samples.

Matching of spectroscopically and X-ray selected groups requires first an
identification of the X-ray group redshift and then a full redshift space
match between the two samples of groups. Although both the redshift of X-ray
and spectroscopically selected groups are derived using galaxy redshifts,
there are distinct differences in these two procedures, allowing us to treat
the X-ray and spectroscopically selected group samples as almost
independent.  X-ray emission is proportional to density squared and is
mainly detectable from cores of galaxy groups, occupying $\sim$10\% of the
total group area. Such a small area has typically only a handful of galaxies
and the goal of targeted spectroscopic identification is to go deeper in
this region and to increase the completeness.  Detection of spectroscopic
groups on the other hand is entirely based on the depths and sampling of the
spectroscopic survey, and typically uses galaxies at much larger separations
compared to the size of the X-ray detection.

\begin{figure*}
\includegraphics[width=18.cm,clip]{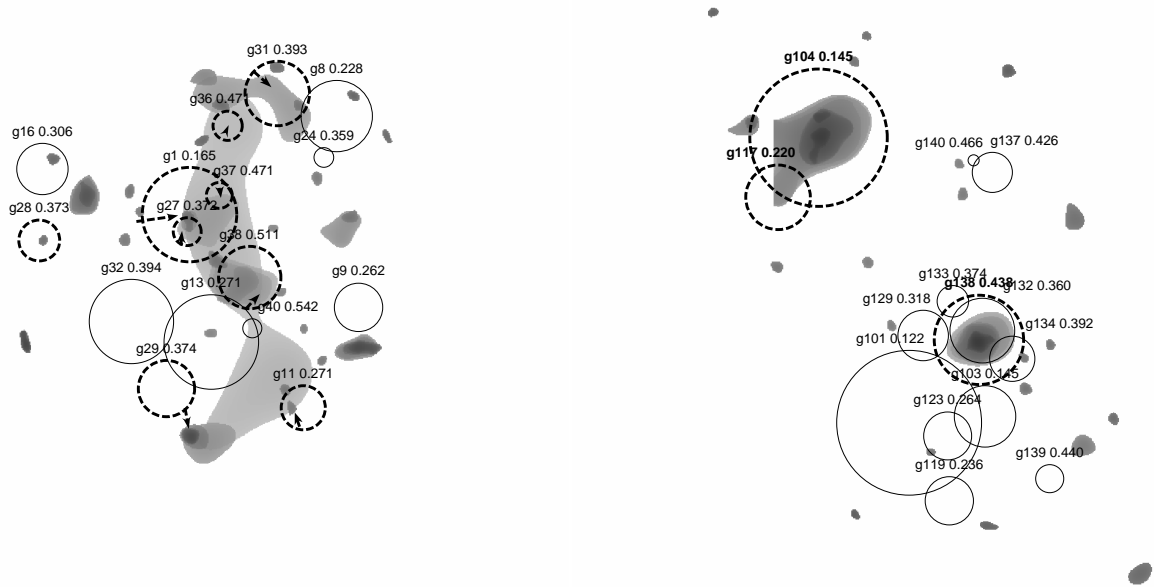}
\centering \figcaption{Wavelet reconstruction of the X-ray surface
  brightness in the 0.5--2 keV band for the RA14h (left) and the RA21h (right)
  CNOC2 patches. Circles indicate the position and size ($R_{200}$) of the
  spectroscopically selected groups.  Groups for which we were able to
  ascribe X-ray emission are shown as thick dashed circles. Groups without
  X-ray emission are shown as thin solid circles. In complex situations,
  arrows are used to make an assignment clear. Numerics refer to the
  spectroscopically selected group ID (from Carlberg et al. 2001) and
  corresponding redshift. Some of the assignment of X-ray emission to an
  optical group is tentative and is therefore not listed in Tab.1.}
\end{figure*}

Fig.5 shows our best attempt to assign X-ray emission to the
spectroscopically selected groups in the CNOC2 catalog (ids from Carlberg et
al, 2001).  In this case, the spectroscopic identification of the X-ray
group was relaxed so that only two galaxies of matching redshift are
required inside the area of X-ray emission.  Circles at the position of
spectroscopically selected groups indicate the area enclosed by $R_{200}$
(as computed by Balogh et al, 2007).  Overlapping X-ray sources (confusion)
are common within areas of this size, particularly in the deeper RA14h field.
This can lead to confused association of groups and X-ray sources, as the
distances between groups of all types can be as low or lower than
uncertainties on their centroids.  X-ray centers are good to 10-30\arcsec\
while the luminosity weighted centers of spectroscopically selected groups
rely strongly on membership allocation and especially redshift completeness,
and while a median off-centering value is $\sim$15\arcsec\, a factor of 3
larger deviations are also predicted (Wilman et al., 2009).  This is less of
a problem in the shallower RA21h field within which only highly significant
and well separated groups are matched (20\% of spectroscopic groups in the
X-ray survey area).  In contrast, the deep RA14h field has a substantially
higher fraction of probable matches (56\% of spectroscopic groups in the
X-ray survey area), but confusion of sources and centering uncertainty
sometimes lead to ambiguous assignation of groups.  The largest offset seems
to be an edge effect -- only part of the rich galaxy group g29 entered the
CNOC2 spectroscopic survey area.  Allocated sources are indicated in Fig.5,
and include probable identifications of confused X-ray sources.  However we
do not exclude the possibility that some associations are missed due to
confused X-ray emission or as yet unidentified (in redshift) X-ray peaks.

We now discuss all cases of tentative assignment in the RA14h field
individually.  Two spectroscopically selected groups (ids g36, g37 at
z=0.471) in the RA14h field are found in an area of very faint and extended
X-ray emission. Detailed analysis shows a low surface brightness 4\arcmin\
scale X-ray detection which corresponds to a highly extended and elongated
collection of galaxies at $z=0.471$ linking the two groups.  X-ray emission
from each of these groups is too faint for individual detection, but
corresponds to approximately $M_{200}\sim 2\times10^{13}M_\odot$, assuming
an equality of their contribution to the detected flux. Emission strictly
associated with group g31 is marginal (1 $\sigma$ significance) and
corresponds to a group mass of $1.4\pm0.7 \times 10^{13} M_\odot$
($M_{200}$). Emission from the group g38 at $z=0.511$ is confused with the
foreground group at $z=0.04$ (see \S\ref{C}).  The X-ray emission present on
1\arcmin\ scales (see Fig.5) can only be explained by a 0.04 group, with
$M_{200}= 7.3\pm0.6\times10^{12}M_\odot$.  However, enhanced X-ray emission
centered on g38 and a high density of galaxies at $z\sim0.511$ coincident
with this emission suggests that g38 can still contribute on smaller scales
with an upper limit on mass of $M_{200}< 6\times10^{13}M_\odot$.  The group
g27 is cospatial with the group g01, but there is a small-scale elongation
of X-ray emission coincident with the location of the core of g27 group. A
corresponding mass estimate for g27 associated with this emission yields
$M_{200}= 2.4\pm0.5 \times10^{13}M_\odot$. Of the groups g27, g31, g36, g37
and g38 none meets the criteria for inclusion into a table of X-ray groups.
However all are likely emitters, and they are located in the zones with
detected X-ray flux.  Excluding these groups would lead to a lower limit of
25\% (4/16) X-ray detections for spectroscopically selected groups in the
RA14h field.

\includegraphics[width=9.cm,clip]{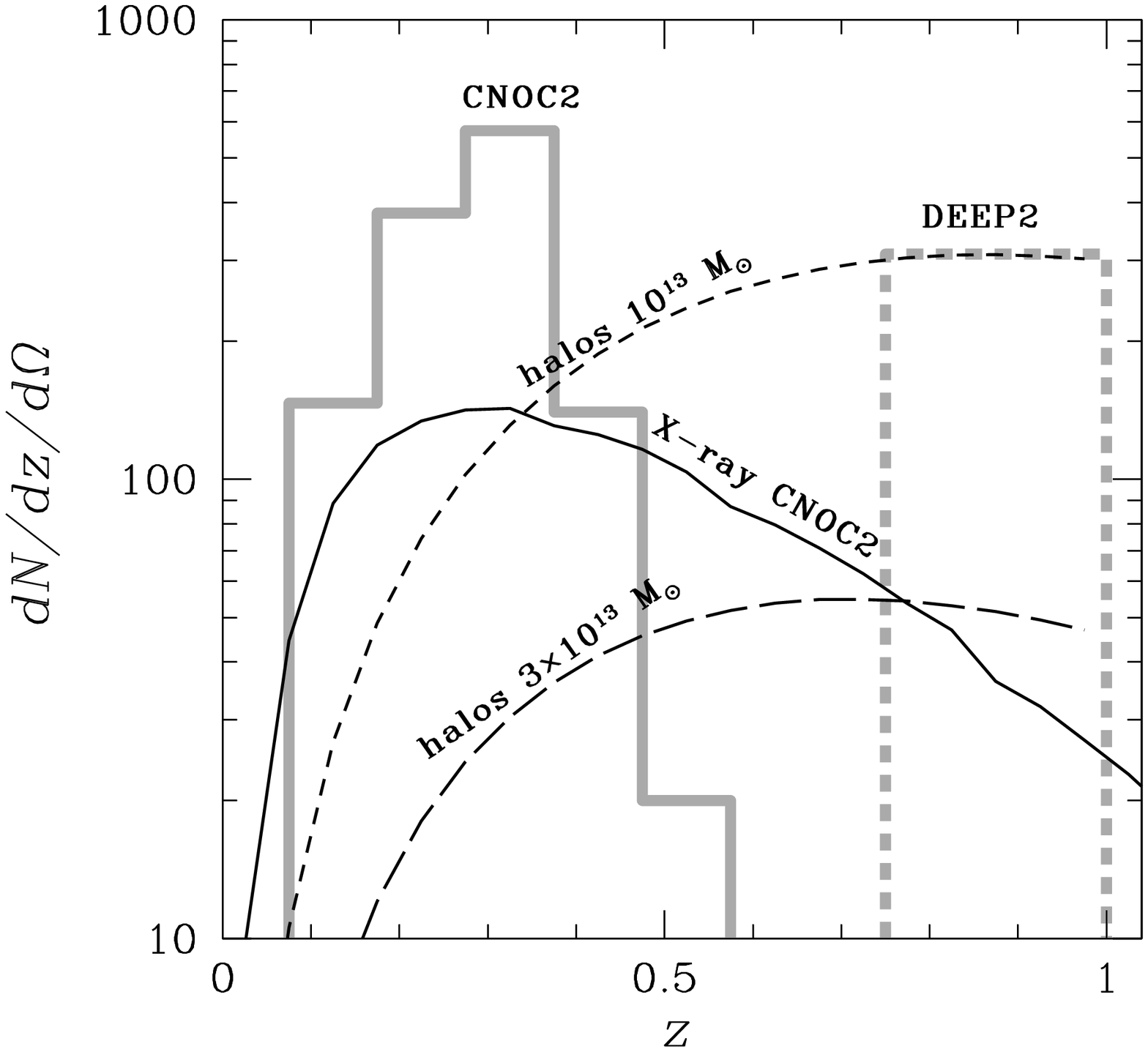} {\centering \figcaption{
    Predicted number density as a function of redshift, dN/dz/d$\Omega$
    (dN/dz per unit area in deg$^{2}$) of galaxy groups.  The solid curve
    shows the prediction for systems found in the X-ray survey.  Short
    (long) dashed line shows the prediction for $10^{13}M_\odot$
    ($3\times10^{13}M_\odot$) halos in WMAP5 cosmology.  Gray histograms
    show the observed abundance of spectroscopically selected groups found
    in CNOC2 (solid) and DEEP2 (dashed) surveys.}}

Fig.6 presents a model of the redshift distribution of X-ray selected
systems (solid line), adding the area and sensitivity of the two CNOC2
patches.  This assumes a WMAP5 cosmology (Komatsu et al. 2008) and the
$L_X-M$ relation from Rykoff et al. (2008) with evolutionary corrections
discussed above.  The short (long) dashed line illustrates the number
density of halos of mass $\geq 10^{13}M_\odot$ ($3\times10^{13}M_\odot$),
and demonstrates the relative contribution of halos above these mass
thresholds.  The X-ray detection mass threshold (where the solid and dashed
lines cross) increases with redshift, from below $10^{13} M_\odot$ at
$z\lesssim0.3$ to above $3\times 10^{13} M_\odot$ at $z\gtrsim0.8$.  X-ray
selection at this depth provides groups at $0\lesssim z\lesssim1$, and so
although the total number of systems per square degree is high ($\sim$100),
the expected match to spectroscopically selected groups within a more
limited redshift interval is moderate.  Nonetheless, the depth of our
observations provides a peak in the redshift distribution at $0.2\lesssim z
\lesssim0.7$, which is well suited to the CNOC2 redshift range of
$0.1\lesssim z \lesssim0.55$. The solid gray histogram, illustrating the
averaged number density of CNOC2 spectroscopically selected groups, contains
roughly twice the number of X-ray groups within this redshift range. Thus, a
naive estimate of X-ray detected groups would be $\sim50\%$; however, we
note that there is a strong variation in the efficiency of group detection in
the CNOC2 survey that is not included here.  Nonetheless, this effectively
demonstrates that our X-ray survey provides a selection of groups down to a
canonical mass value of $10^{13} M_\odot$, below which the use of X-ray
selection is not yet established.

This contrasts with the situation at higher redshift, for which the number
density of X-ray detected groups is expected to drop off as the detection
threshold gets pushed to higher and higher mass.  This is the case for the
DEEP2 spectroscopically selected group sample (I) at $0.75<z<1.03$ from
Gerke et al. (2007, dashed gray histogram), which shows that at exposures
similar to our survey the full strength of the X-ray selection of galaxy
groups is not yet exploited, while the DEEP2 spectroscopic galaxy group
survey would allow a comparison down to the $10^{13} M_\odot$ mass limit.

The modeling in Fig.6 makes a simplistic assumption that the spectroscopic
survey for galaxy groups is complete, which may introduce further
differences between spectroscopic and X-ray selected group samples, which we
consider next.

\newpage

\includegraphics[width=9.cm,clip]{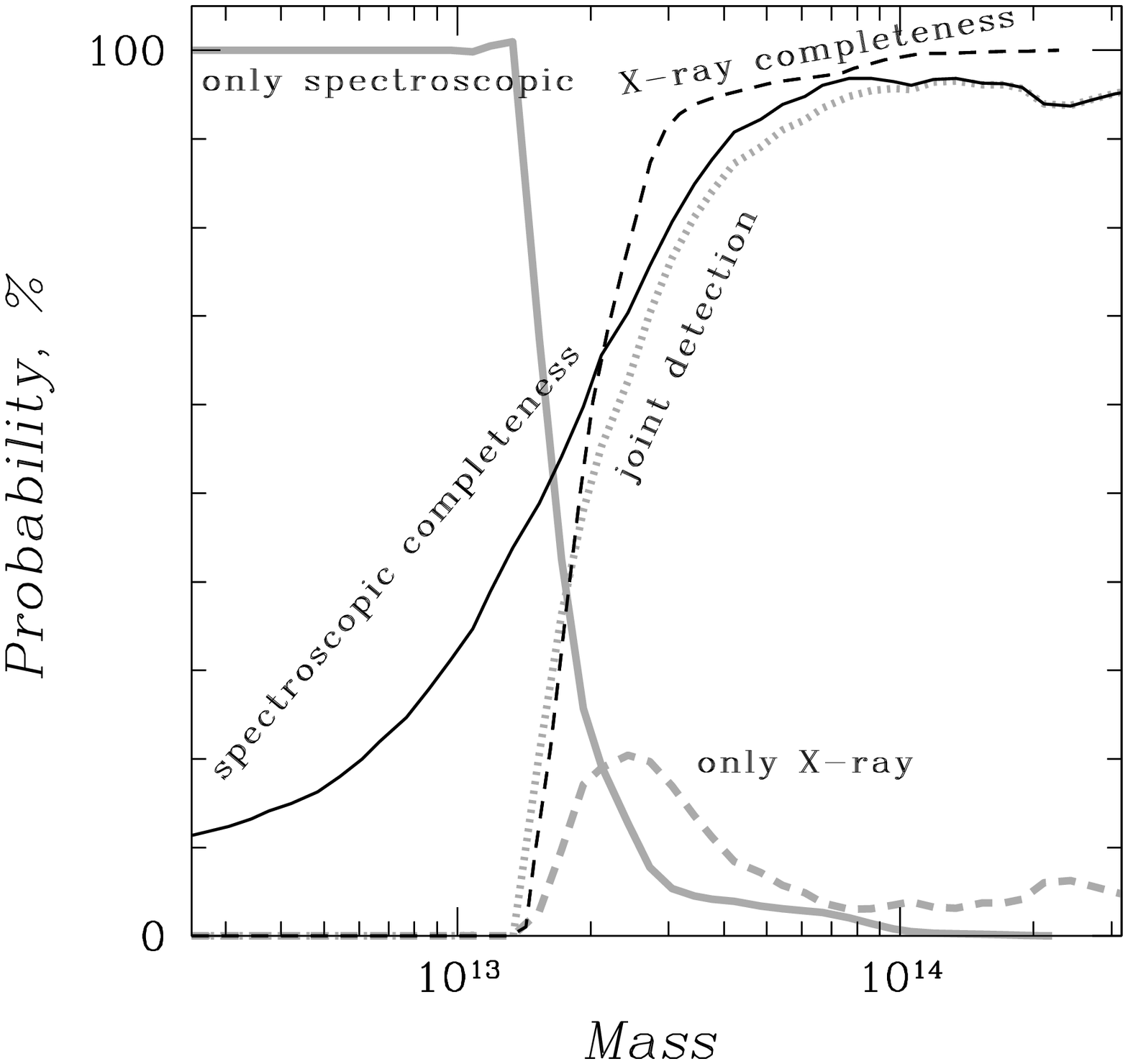} {\centering
  \figcaption{Probability of group detection as a function of halo mass at
    $z\sim0.4$.  Both X-ray selection at the depth of the RA14h field (dashed
    black curve) and spectroscopic selection in CNOC2 (including the effects
    of incompleteness, solid black curve) are modeled as described in the
    text.  We ignored a possible covariance between optical richness and
    X-ray luminosity in predicting the percentage of groups only detected in
    X-rays (dashed gray line), the percentage of groups only detected
    spectroscopically (solid gray line) and the percentage of groups
    detected jointly (dotted gray line). }}

Fig.7 illustrates the recovery of groups as a function of halo mass,
detected using both X-ray (at RA14h depth) and spectroscopic selection
methods.  The figure is constructed to show the percentage of groups
detected as a function of mass, evaluated at a typical CNOC2 redshift
$z=0.4$.  X-ray groups are modeled as in Fig.6, and scatter in the Lx-M
relation is ignored, providing a mass threshold which is a simple function
of halo mass: the non-abrupt mass cut-off is merely a result of the variable
depth across the RA14h field.  Introducing scatter would introduce a higher
sensitivity toward low-mass systems, smearing the boundary by additional
30\% in mass (Vikhlinin et al. 2009). The modeling of the group recovery
rate in the CNOC2 spectroscopic survey is based on applying the
spectroscopic survey characteristics (including mean sampling rate) to the
semi-analytic galaxy formation model in Millennium Simulation (Font et al.
2008), as described by McGee et al. (2008).  Spectroscopically selected
samples will inevitably include a large number of lower mass groups, as the
drop in recovery rate is compensated by an increase in number density.  The
correlation of the richness and X-ray luminosity is accounted in the plot,
while we ignored the second-order effects associated with the possible
covariance in the deviation from the richness-mass and Lx-mass relations.

It is interesting to examine the expected mass of groups detected only by
X-ray or spectroscopic methods, and the intersection between the two samples
(indicated by gray lines in Fig.7) to complement the actual data.  In the
RA14h field, all three of the confirmed $0.14\leq z\leq0.5$ X-ray selected
groups are also in the spectroscopically selected sample (which is only
sensitive to this limited redshift range).  In the shallower RA21h field only
three out of the ten $0.14\leq z\leq0.5$ significant X-ray groups within the
area covered by the CNOC2 survey are in the spectroscopically selected
sample. Of the seven undetected groups, five have estimated masses less than
$\sim 3\times10^{13}M_{\odot}$; from Fig.7 we see that the spectroscopic
completeness in this mass range is less than $\lesssim 80$ per cent, due
primarily to the sparse sampling; therefore perfect recovery of such
low-mass X-ray groups is not expected.  However, two X-ray groups with
masses $>7\times10^{13}M_{\odot}$ are absent from the spectroscopic
catalog.  This may be due to the spatial variation of the spectroscopic
completeness, which is not included in the completeness function of Fig. 7;
we also note that group X21h13 is right near the edge of the CNOC2
coverage.
Thus, overall we find 6/13 X--ray detected groups are also identified in an
independent spectroscopic survey, over the area and redshift range that they
can be fairly compared.  This is approximately consistent with expectations,
given the sparse sampling of that survey.  Only one group is a surprising
non-detection, given its estimated mass.  Better sampled spectroscopy and
improved mass estimates for these systems will help to understand this
anomaly.  But we find no convincing evidence for a substantial population of
X--ray bright groups with no optical counterpart.

\includegraphics[width=9.cm,clip]{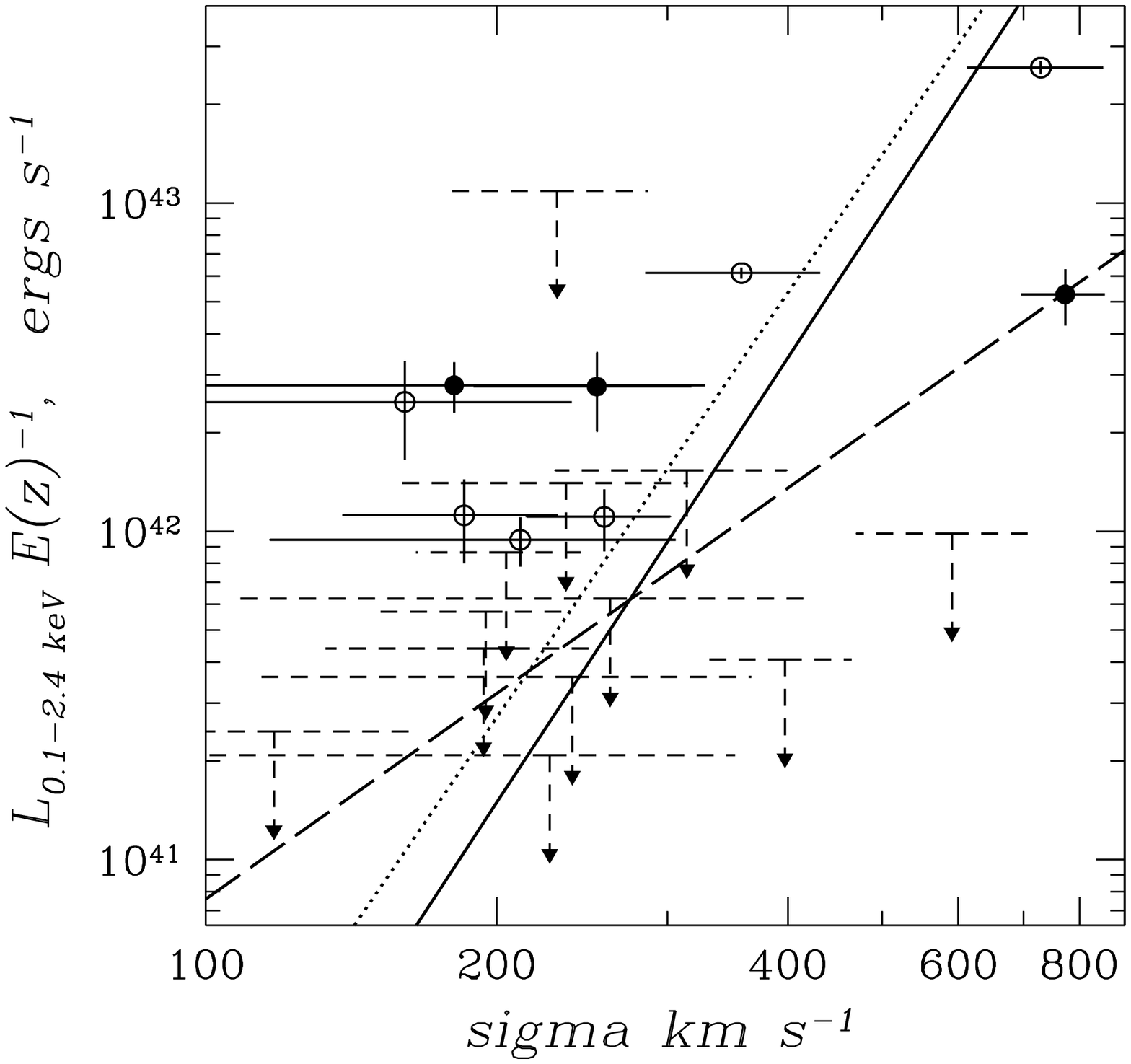} {\centering
  \figcaption{ $L_X-\sigma$ relation for groups. The open points are our
    measurements of the flux attributed to the uniquely identified groups.
    The filled points correspond to detections of emission at the position
    of optical groups, but which are likely contaminated with emission from
    other groups. The $2\sigma$ upper limits are shown using short-dashed
    line ending with an arrow. 68\% error bars are shown for the velocity
    dispersion measurements. In calculating the X-ray luminosity, the
    K-correction has been done iteratively using both the redshift of the
    source and the expected temperature given the evolved L--T relation (see
    Finoguenov et al. 2007 for details). The fits to the local sample data
    of Mulchaey et al. (2003), as presented in Spiegel et al., are shown as
    solid (inverse regression) and dashed (direct regression) lines. The fit
    to the $z\sim 0.25$ data of Rykoff et al.  (2008) is shown as the dotted
    line. All fits were corrected for differences in the definition of
    energy band for calculating $L_X$ and evolution. We find no evidence for
    differences in the $L_X-\sigma$ relation for low and high redshift
    groups.}}

We now address the flip-side of this issue, on the X--ray detection of
groups selected from optical spectroscopy.  There has been recent discussion
in the literature on the lack of detection of X-ray emission from optical
groups at high redshifts (Spiegel et al. 2007; Fang et al. 2007).  In
particular, Spiegel et al. (2007) claim that their upper limits on the X-ray
emission from the groups in the CNOC2 RA14h patch are mildly inconsistent
with the optical-X-ray scaling relations found for the low redshift groups
in Mulchaey et al. (2003).  By combining X-ray data from Chandra and
XMM-Newton, we are able to detect several groups which were not detected by
Spiegel et al. (2007); furthermore, from our follow-up spectroscopy we have
better determined velocity dispersion measurements (Wilman et al. 2005a).
Based on these new measurements, we present in Fig.8 the X--ray luminosities
and limits as a function of group velocity dispersion (the equivalent of
Spiegel et al.'s Fig.4), for both the RA14h and RA21h patches. The evolved
local relation is shown as the solid and dashed lines; these are based on
the data of Mulchaey et al. (2003), as fit by Spiegel et al.; the two fits
correspond to using either direct or inverse regression.  We also show the
best-fit relation of the $z\sim 0.25$ sample from Rykoff et al. (2008).  All
of our detections lie near or above these best-fit relations, and within the
scatter defined by the Mulchaey et al. (2003) data.  Moreover, all but two
of our upper limits are consistent with these relations and, again, fully
consistent within the scatter of those local data.  Therefore we find no
evidence for a significant difference between the X-ray properties of
moderate redshift groups and local samples.

Moreover, we would like to outline a number of caveats
that must be considered when comparing the CNOC2 groups with local samples.
First, Mulchaey et al. (2003) was based on an archival ROSAT sample.  Such
archival samples tend to be dominated by X-ray bright systems (see
discussion in Mulchaey 2000), which introduces a bias against high velocity
dispersion, low $L_X$ groups. Therefore, a comparison between the
optically-selected CNOC2 groups and the low redshift sample in Mulchaey et
al. (2003) is not straightforward.  In addition, the scatter in the
$L_X-\sigma$ relation is large and requires as large a scatter in the
$\sigma-M$ relation as in the $L_X-M$ relation, as confirmed in the analysis
of numerical simulations (Biviano et al. 2006). Such an intrinsic scatter
reduces the significance of deviations from the best-fit relation of any
subsample, which was not taken into account in Spiegel et al. (2007).  
Thus, we find no evidence that high redshift groups are anomalously faint,
as they would be if the gas had been subject to very strong preheating or
feedback (e.g., Balogh et al. 2006). No significant evidence for systematic
differences in X-ray properties between nearby and moderate-z groups is
consistent with the results for the X-ray selected samples discussed by
Jeltema et al. (2006).

\section{Summary}\label{S}

We have presented an X-ray survey of two CNOC2 fields with Chandra and
XMM-Newton and our method to find extended X-ray emission.  We described
preliminary results from the spectroscopic follow-up of the X-ray selected
groups and provide a catalog of X-ray group candidates.  To match
spectroscopically and X-ray selected groups it is crucial to provide a full
3D match in redshift space.  Centering issues mean a purely 2D match will
always be insufficient, especially given incompleteness of spectroscopic
surveys and the fact that current X-ray observations are only able to detect
the central regions of groups.  A probable detection of $\sim50\%$ ($20\%$)
of spectroscopically selected groups in the deeper (shallower) RA14h (RA21h)
field demonstrates that a depth of $\geq$300 ksec with Chandra is crucial to
reach the sensitivity necessary for X-ray detection of galaxy groups in the
redshift range $0.1\lesssim z\lesssim0.6$.  The X-ray groups with identified
spectroscopic redshifts have an ensemble-averaged weak lensing velocity
dispersion of 309$\pm106$ km s$^{-1}$. Finally we show that our current data
show no statistically significant evidence for any mismatch between the
X-ray and spectroscopically selected groups. However improved statistics and
mass estimates, which we will have once on-going XMM, spectroscopic and
multiwavelength datasets are complete and evaluated, will facilitate better
comparison of X-ray and optical properties of groups and allow us to test
the origin of scatter and evolution in the $L_X-M$ relation.

\acknowledgments

Support for this work was provided by the National Aeronautics and Space
Administration through Chandra Award Number G06-7090X issued by the Chandra
X-ray Observatory Center and through NASA grants NNG06AG02G and NNG04GC846.
Based on observations made with ESO Telescopes at Paranal under program ID
080.A-0427 and 081.A-0103 and the Magellan telescopes operated by The
Carnegie Institution of Washington. We acknowledge the use of the FORS and
COSMOS pipelines.  AF acknowledges support from BMBF/DLR under grants
50OR0207 and 50OR0204 to MPE. LCP and MLB acknowledge support from a NSERC
Discovery Grants. The authors thank MPE and the University of Waterloo for
the hospitality during the group meetings held in 2007 and 2008. The authors
thank Mike Hudson for his contribution to the project. We thank the CNOC2
team for the use of their unpublished redshifts and the anonymous referee for
detailed comments which improved the presentation aspects of the paper.  AF
thanks Gabriel Pratt for useful comments on the manuscript.

\end{document}